**Flexible Transparent Field-Effect Diodes Fabricated at Low-Temperature with All Oxide Materials**

*Yonghui Zhang, Zengxia Mei,\* Shujuan Cui, Huili Liang, Yaoping Liu and Xiaolong Du*

Y. H. Zhang, Prof. Z. X. Mei, S. J. Cui, Dr. H. L. Liang, Prof. Y. P. Liu, Prof. X. L. Du
Beijing National Laboratory for Condensed Matter Physics, Institute of Physics, Chinese Academy of Sciences, Beijing 100190, P. R. China
E-mail: zxmei@aphy.iphy.ac.cn

Flexible and transparent electronics[1–3] have gained momentum in the last decade, owing to their great potential in future technological applications.[4–6] Most of the researches, concerning about inorganic semiconductors, have been generally involved in metal oxide semiconductors, as they are optically transparent and compatible with low temperature process. Flexible and transparent oxide thin-film transistors (TFTs), as key devices for realizing next generation circuits, have been extensively studied both on glass and plastic substrates during these years.[7–9] High performances with field-effect mobility more than 10 cm$^2$ V$^{-1}$ s$^{-1}$, on/off current ratio more than 10$^8$ and subthreshold swing less than 0.5 V dec$^{-1}$ were obtained.[7–9] Besides TFTs, thin-film diodes are important components to achieve electronic circuits, especially for energy conversion[10,11] and selective switching.[12,13]

However, few researches have focused on flexible or transparent diodes. The limited reports can be categorized into 4 types: (1) *pn* heterojunction diode.[14–19] As most of the wide-bandgap semiconductors are n-type conductive, a proper p-type wide-bandgap material must be chosen wisely to form a large built-in potential barrier. (2) Schottky junction diode.[20–23] The electron affinities ($\chi$) of most wide-bandgap materials are more than 4 eV, thus only a small Schottky barrier height (*SBH*) could be formed with non-noble metals. (3) Metal-insulate-semiconductor (MIS) diode.[24] As in Schottky diode, a large difference between metal work function ($\Phi_M$) and semiconductor affinity ($\chi$) is needed to achieve a large rectification ratio. (4) Metal-insulator-metal (MIM) diode.[25,26] Actually, it is not easy for MIM diodes to be applied in transparent circuits because of the difficulties to find two kinds of transparent electrodes with large work function difference. (5) Self-switching diode (SSD).[27] Besides small rectification ratio, this kind of device involves nanofabrication, which may bring high costs and challenges in technology compatibility.

Diode-connected field-effect transistor (FET)/bipolar junction transistor (BJT), with drain/collector and gate/base electrodes shorted, is widely used in integrated circuits to server as passive load.[28] Because of its asymetric current-voltage characteristics,[29,30] diode-connected transistor is also used as rectification device in energy harvest systems.[31,32] This paper reports, to the best of our knowledge, the first flexible and fully transparent field-effect diode (FED) fabricated at low-temperature with all oxide materials, using diode-connected thin-film transistor architecture. Conventional thin-film transistors were also fabricated at the same time as reference. The diodes exhibited a high rectification ratio of $5 \times 10^8$ and low leakage current of 1 pA, same as the on/off ratio and off current in the referenced TFT. Technology computer aided design (TCAD) simulation was employed to explore its working mechanism. Finally, a single-stage rectifier was demonstrated by applying this unique FED to rectify alternating current (AC) signals with different amplitudes and frequencies.

For ease of understanding, a conceptualized schematic of FED is shown in **Figure 1**a. The device is two-terminal contacted, with dielectric and channel layers stacked inside. A low-temperature ($\leq 100$ °C) process (Figure 1b) was used to fabricate the devices. (Details of the fabrication process can be found in Experimental Section.) During the same process, a conventional TFT was fabricated as well to serve as reference (see the top view in Figure 1c). Both of the devices on polyethylene naphthalate (PEN) and glass substrates are optically transparent, with the whole devices (including the substrates) exhibiting a transmittance over 80% in full visible spectral range (Figure 1d).

The referenced TFT adopted a staggered bottom-gate structure, with channel width/length (*W/L*) of 300/20 μm. **Figure 2**a shows the transfer characteristics of referenced TFT on PEN substrates ($I_{DS}$-$V_{GS}$, DS denotes drain to source and GS gate to source). An on/off current ratio of 5 × 10^8 was obtained by operating the TFT at $V_{DS}$ = 3 V and $V_{GS}$ = ± 20 V. The field-effect mobility ($\mu_{eff}$) of 11.56 cm$^2$ V$^{-1}$ s$^{-1}$ was calculated from the linear region. The subthreshold swing (*SS*) was determined to be 0.53 V dec$^{-1}$, calculated using the minimum value of $1/(\partial \log(I_{DS})/\partial V_{GS})$ versus $V_{GS}$ plot. The turn-on voltage ($V_{ON}$) was read out to be ~ 0 V from the transfer curve. Gate oxide capacitance ($C_{ox}$) of $1.5 \times 10^{-7}$ F/cm$^2$ was extracted from capacitance-voltage (*C-V*) measurement curve (not shown here). Figure 2b shows the $I_{DS}$-$V_{DS}$ output characteristics at $V_{GS}$ = 5 - 10 V, which exhibiting typical square-law behavior. It should be noted that good ohmic contact was formed at the interface of indium tin oxide/zinc oxide (ITO/ZnO), so no rectifying property was expected from source and drain contacts. These properties in referenced TFT are important, as the field-effect diode has the identical material quality and will follow the same field-effect principles. Devices on glass substrates have similar characteristics with those on PEN, so, to avoid repetition, the electrical characteristics in this paper are all carried out on the PEN devices.

Figure 2c shows the current-voltage (*I-V*) characteristics of FED in flat and bent states, respectively (see Figure 2d). A high rectification ratio of 5 × 10^8 was obtained at *V* = ± 20 V, which is much higher than most of the reported Schottky and *pn* junction diodes.[14,15,18–20,33–35] The current of bent-test was one order of magnitude smaller than those of flat-test, which may be caused by the bad contact between the probe tips and the electrodes while the device was bent. Besides that, there was minimal change in device performance when the substrate was bent up to *r* = 8 mm. This indicates that the presented oxide diode has large rectification ratio and good robustness, which makes it a promising candidate for flexible and transparent electronics.

To illustrate the working principle of the FED, simulations were carried out by ATLAS simulator, included in the Silvaco TCAD software, to visualize the electrical characteristics under different voltage biases. A simulated *I-V* characteristic is shown in **Figure 3**a, with the defined device structure in inset. Despite the minor rectification ratio of 10^4, a similar rectifying behavior appears. Figure 3b shows the conduction band energy ($E_C$) distribution of FED at *V* = -1 V (Figure 3b i) and $V = 1$ V (Figure 3b ii). Along cutline, $E_C$ increases monotonically when $V = -1$ V (Figure 3b iii), while decreases when $V = 1$ V (Figure 3b iv). As a consequence, the electron concentration ($Ne$), near ZnO/Al$_2$O$_3$ interface, was cut down when $V = -1$ V (Figure 3c i), while boosted up when $V = 1$ V (Figure 3c ii). Eventually, no obvious current formed between anode and cathode when $V = -1$ V (Figure 3d i), while clear current path formed when $V = 1$ V (Figure 3d ii). Clearly, this is where the rectifying behavior comes from. As can be seen in the insets of Figure 2a & 2c, the main difference between diode-connected and conventional TFT devices is that it has gate and drain shorted. Actually, the shorted gate-drain electrode is the anode in FED, and it both served as gate, to form conductive channel, and drain, to collect channel electrons, at the same time.

Furthermore, a single-stage rectifier was built and characterized to demonstrate the applicability of this diode. As shown in **Figure 4**a, a Keithley 3390 arbitrary waveform generator was employed to input AC signals, in series with FED and a 15 MΩ resistor. A Tektronix TDS1000B-SC oscilloscope was in parallel connection with the resistor to measure the output voltage. AC sine signals with amplitude among 5 V - 10 V, frequency of 1 Hz, were input into the circuit (Figure 4b i), and half wave rectification was observed (Figure 4b ii). The output voltages were smaller than the input voltages, which was partly caused by the big threshold voltage ($V_{TH}$) and subthreshold swing ($SS$). Besides, the output voltage was measured across the 15 MΩ resistor, which was comparable with the channel resistance, thus only a fraction of $V_{IN}$ dropped on FED. Figure 4c shows the half wave rectification of AC square signals with frequency among 1 Hz - 1 MHz at amplitude of 10 V. Signals with frequency less than 1 KHz had distinguishable waveform while signals with frequency between 10 KHz and 100 KHz didn't follow the waveform well. For signals with frequency above 1 MHz, no rectification was observed. The cut-off frequency of ~ 1 MHz in this paper is lower than some other reported *pn* and Schottky diodes.[18,21] This may be attributed to the poor crystalline quality of ZnO, as it was deposited by sputtering at room temperature. Besides, the big device size results in big capacitance, which also leads to a low cut-off frequency. Nevertheless, like Schottky diodes, FED works via majorities, so it is expected to be operated in high frequency up to GHz.[36]

In summary, we have reported flexible fully transparent field-effect diodes, fabricated at low temperature (≤100 °C) with all oxide materials. The device is optically transparent with transmittance over 80% in visible spectra range and robust while mechanically bending up to $r = 8$ mm. Distinguished from other junction diodes, this diode follows field-effect principle as TCAD simulation reveals. The cathode serves as source, to eject electrons, while the anode both served as gate, to form conductive channel, and drain, to collect channel electrons, at the same time. The diodes exhibit high rectification ratio of $5 \times 10^8$ and low leakage current of 1 pA. Half wave rectification was achieved by a single-stage rectifier with a cut-off frequency of 1 MHz. To further increase the device performance, more investigations could be done to optimize the threshold voltage, subthreshold swing, crystalline quality and device geometry.

**Experimental Section**

*Device fabrication*: Field-effect diodes and referenced TFTs were fabricated on quartz glass (500 μm thick) and Polyethylene naphthalate (PEN) substrates (125 μm thick). Figure 1b shows the experimental scheme for the fabrication of field-effect diodes. Step 1: PEN and glass substrates were ultrasonic cleaned in acetone and isopropyl alcohol and blown dry with nitrogen. Step 2: A bottom indium tin oxide (ITO) anode (200 nm thick) was deposited by radio frequency (RF) magnetron sputtering and patterned by UV-lithography via lift-off. Step 3: A $Al_2O_3$ dielectric layer (50 nm thick) was deposited by atomic layer deposition (ALD) at a maximum temperature of 100 °C and patterned by UV-lithography followed by wet etching in phosphoric acid ($H_3PO_4$ 90% for 90 s at 50 °C). Step 4: A ZnO active layer (50 nm thick) was deposited by RF-sputtering with a ceramic target (99.99%) at room temperature and patterned by UV-lithography followed by wet etching in hydrochloric acid (HCl 1% for 3 s). Step 5: The same ITO sputtering and lift-off process was used to define the top contacts.

*Device characterization*: Electrical characteristics measurements in Figure 2 were measured in air using source-measurement unit (SMU) and capacitance-voltage unit (CVU) included in Keithley 4200 semiconductor characterization system. The field-effect mobility in the linear region was calculated as a function of gate voltage using $\mu_{eff} = (W/L) \times G_m/(C_{ox} \times V_{DS})$. $G_m$ is transconductance extracted from the transfer curve using $G_m = \partial I_{DS}/\partial V_{GS}$.

*Device simulation*: The simulations in Figure 3 were performed by ATLAS simulator included in the Silvaco TCAD software. A non-linear mesh was defined to accurately characterize in active areas while coarsely elsewhere. On this basis, a ZnO layer (50 nm thick, energy gap $E_G = 3.37$ eV, affinity $\chi = 4.5$ eV) and a $Al_2O_3$ layer (50 nm thick, dielectric constant $\varepsilon_r = 9.3$) were defined to serve as channel and dielectric layer, with conductor (work function $\Phi_M = 4.33$ eV) severing as contact electrodes. Fermi-Dirac statistics model was used to get a precise description of electrons in thermal equilibrium.

**Acknowledgements**

This work is supported by the grants from the Ministry of Science and Technology of China (Grant Nos. 2011CB302002, and 2011CB302006), the National Science Foundation of China (Grant Nos. 11274366, 51272280, 11174348, 61204067, and 61306011), and the Chinese Academy of Sciences.

Received: ((will be filled in by the editorial staff))
Revised: ((will be filled in by the editorial staff))
Published online: ((will be filled in by the editorial staff))

[1]     K. Nomura, H. Ohta, K. Ueda, T. Kamiya, M. Hirano, H. Hosono, *Science* **2003**, *300*, 1269.

[2]     J. F. Wager, *Science* **2003**, *300*, 1245.

[3]     K. Nomura, H. Ohta, A. Takagi, T. Kamiya, M. Hirano, H. Hosono, *Nature* **2004**, *432*, 488.

[4]     J.-S. Park, T.-W. Kim, D. Stryakhilev, J.-S. Lee, S.-G. An, Y.-S. Pyo, D.-B. Lee, Y. G. Mo, D.-U. Jin, H. K. Chung, *Appl. Phys. Lett.* **2009**, *95*, 013503.

[5]     A. K. Tripathi, E. C. P. Smits, J. B. P. H. van der Putten, M. van Neer, K. Myny, M. Nag, S. Steudel, P. Vicca, K. O'Neill, E. van Veenendaal, J. Genoe, P. Heremans, G. H. Gelinck, *Appl. Phys. Lett.* **2011**, *98*, 162102.

[6]     C. Dagdeviren, S.-W. Hwang, Y. Su, S. Kim, H. Cheng, O. Gur, R. Haney, F. G. Omenetto, Y. Huang, J. A. Rogers, *Small* **2013**, *9*, 3398.

[7]     E. Fortunato, P. Barquinha, R. Martins, *Adv. Mater.* **2012**, *24*, 2945.

[8]     T. Rembert, C. Battaglia, A. Anders, A. Javey, *Adv. Mater.* **2015**, *27*, 6090.

[9]     T. Kamiya, H. Hosono, *Npg Asia Mater.* **2010**, *2*, 15.

[10]    X. Pu, L. Li, H. Song, C. Du, Z. Zhao, C. Jiang, G. Cao, W. Hu, Z. L. Wang, *Adv. Mater.* **2015**, *27*, 2472.

[11]    H. J. Visser, R. J. M. Vullers, *Proc. IEEE* **2013**, *101*, 1410.

[12]    M.-J. Lee, S. Seo, D.-C. Kim, S.-E. Ahn, D. H. Seo, I.-K. Yoo, I.-G. Baek, D.-S. Kim, I.-S. Byun, S.-H. Kim, I.-R. Hwang, J.-S. Kim, S.-H. Jeon, B. H. Park, *Adv. Mater.* **2007**, *19*, 73.

[13]    W. Y. Park, G. H. Kim, J. Y. Seok, K. M. Kim, S. J. Song, M. H. Lee, C. S. Hwang, *Nanotechnology* **2010**, *21*, 195201.

[14]    S. Narushima, H. Mizoguchi, K. Shimizu, K. Ueda, H. Ohta, M. Hirano, T. Kamiya, H. Hosono, *Adv. Mater.* **2003**, *15*, 1409.

[15]    F.-L. Schein, H. von Wenckstern, M. Grundmann, *Appl. Phys. Lett.* **2013**, *102*, 092109.

[16]    F.-L. Schein, M. Winter, T. Böntgen, H. von Wenckstern, M. Grundmann, *Appl. Phys. Lett.* **2014**, *104*, 022104.

[17]    P. Schlupp, F.-L. Schein, H. von Wenckstern, M. Grundmann, *Adv. Electron. Mater.* **2015**, *1*, 1400023.

[18]    W.-C. Chen, P.-C. Hsu, C.-W. Chien, K.-M. Chang, C.-J. Hsu, C.-H. Chang, Wei-Kai Lee, W.-F. Chou, H.-H. Hsieh, C.-C. Wu, J. Phys. *Appl. Phys.* **2014**, *47*, 365101.

[19]    B. N. Pal, J. Sun, B. J. Jung, E. Choi, A. G. Andreou, H. E. Katz, *Adv. Mater.* **2008**, *20*, 1023.

[20]    J. Zhang, Y. Li, B. Zhang, H. Wang, Q. Xin, A. Song, *Nat. Commun.* **2015**, *6*, 7561.

[21]    A. Chasin, M. Nag, A. Bhoolokam, K. Myny, S. Steudel, S. Schols, J. Genoe, G. Gielen, P. Heremans, *IEEE Trans. Electron Devices* **2013**, *60*, 3407.

[22]    L. J. Brillson, Y. Lu, *J. Appl. Phys.* **2011**, *109*, 121301.

[23]    L. J. Brillson, Y. Dong, F. Tuomisto, B. G. Svensson, A. Y. Kuznetsov, D. Doutt, H. L. Mosbacker, G. Cantwell, J. Zhang, J. J. Song, Z.-Q. Fang, D. C. Look, *J. Vac. Sci. Technol. B* **2012**, *30*, 050801.

[24]    T. Sugimura, T. Tsuzuku, Y. Kasai, K. Iiyama, S. Takamiya, *Japanese Journal of Applied Physics* **2000**, *39*, 4521.

[25]    S. Grover, G. Moddel, *IEEE Journal of Photovoltaics* **2011**, *1*, 78.

[26]    S. Hemour, K. Wu, *Proceedings of the IEEE* **2014**, *102*, 1667.

[27]    Y. Kimura, Y. Sun, T. Maemoto, S. Sasa, S. Kasai, M. Inoue, *Jpn. J. Appl. Phys.* **2013**, *52*, 06GE09.

[28]    B. Razavi, *Design of Analog CMOS Integrated Circuits*, Tata McGraw-Hill Education, NY, USA, **2002**.

[29]    W. L. Lim, E. J. Moon, J. W. Freeland, D. J. Meyers, M. Kareev, J. Chakhalian, S. Urazhdin, *Appl. Phys. Lett.* **2012**, *101*, 143111.

[30]    W. Chen, K.-Y. Wong, W. Huang, K. J. Chen, *Appl. Phys. Lett.* **2008**, *92*, 253501.

[31]    Y. Yao, J. Wu, Y. Shi, F. F. Dai, *IEEE Trans. Ind. Electron.* **2009**, *56*, 2317.

[32]    K. Kotani, A. Sasaki, T. Ito, *IEEE J. Solid-State Circuits* **2009**, *44*, 3011.

[33]    Y. C. Shin, J. Song, K. M. Kim, B. J. Choi, S. Choi, H. J. Lee, G. H. Kim, T. Eom, C. S. Hwang, *Appl. Phys. Lett.* **2008**, *92*, 162904.

[34]    H. Ohta, M. Hirano, K. Nakahara, H. Maruta, T. Tanabe, M. Kamiya, T. Kamiya, H. Hosono, *Appl. Phys. Lett.* **2003**, *83*, 1029.

[35]    A. Kudo, H. Yanagi, K. Ueda, H. Hosono, H. Kawazoe, Y. Yano, *Appl. Phys. Lett.* **1999**, *75*, 2851.

[36]    B. Bayraktaroglu, K. Leedy, R. Neidhard, *IEEE Electron Device Lett.* **2009**, *30*, 946.

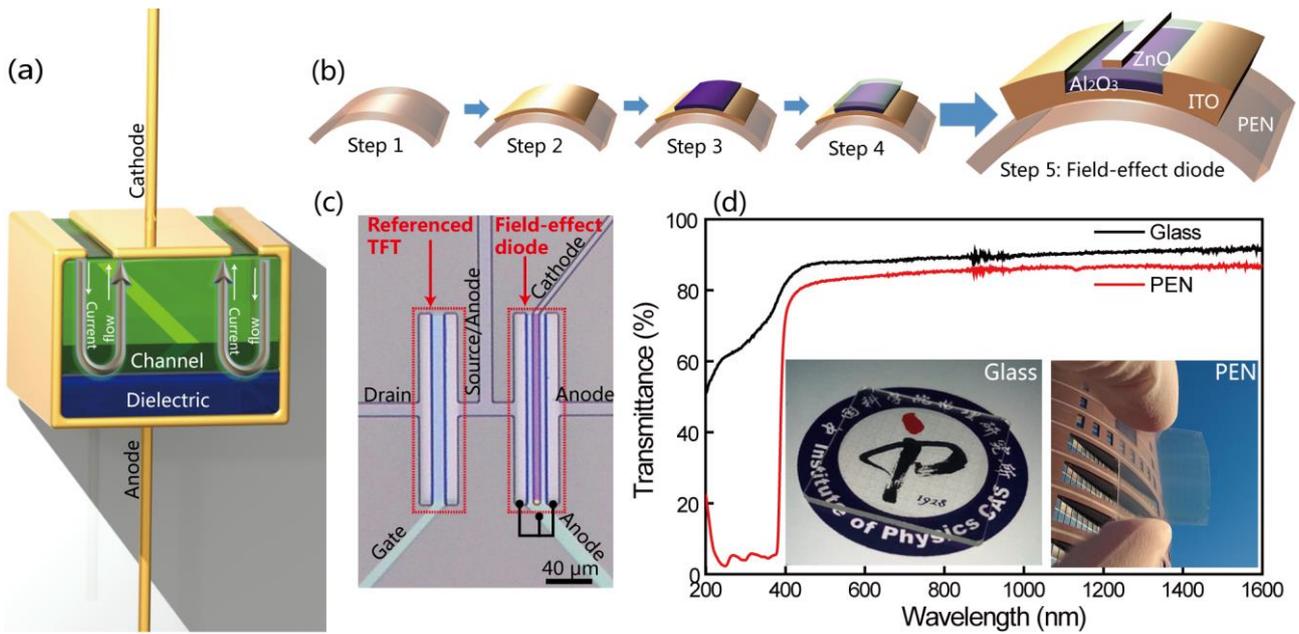

**Figure 1.** Device structure of field-effect diode. (a) Conceptualized structure diagram, which features two-terminal configuration. (b) Fabrication procedure of field-effect diode. (c) Top view of field-effect diode and referenced TFT. (d) Optical transmittance spectra of devices on glass and PEN substrates with transmittance over 80% in visible spectral range. The insets show the photographs of these two devices.

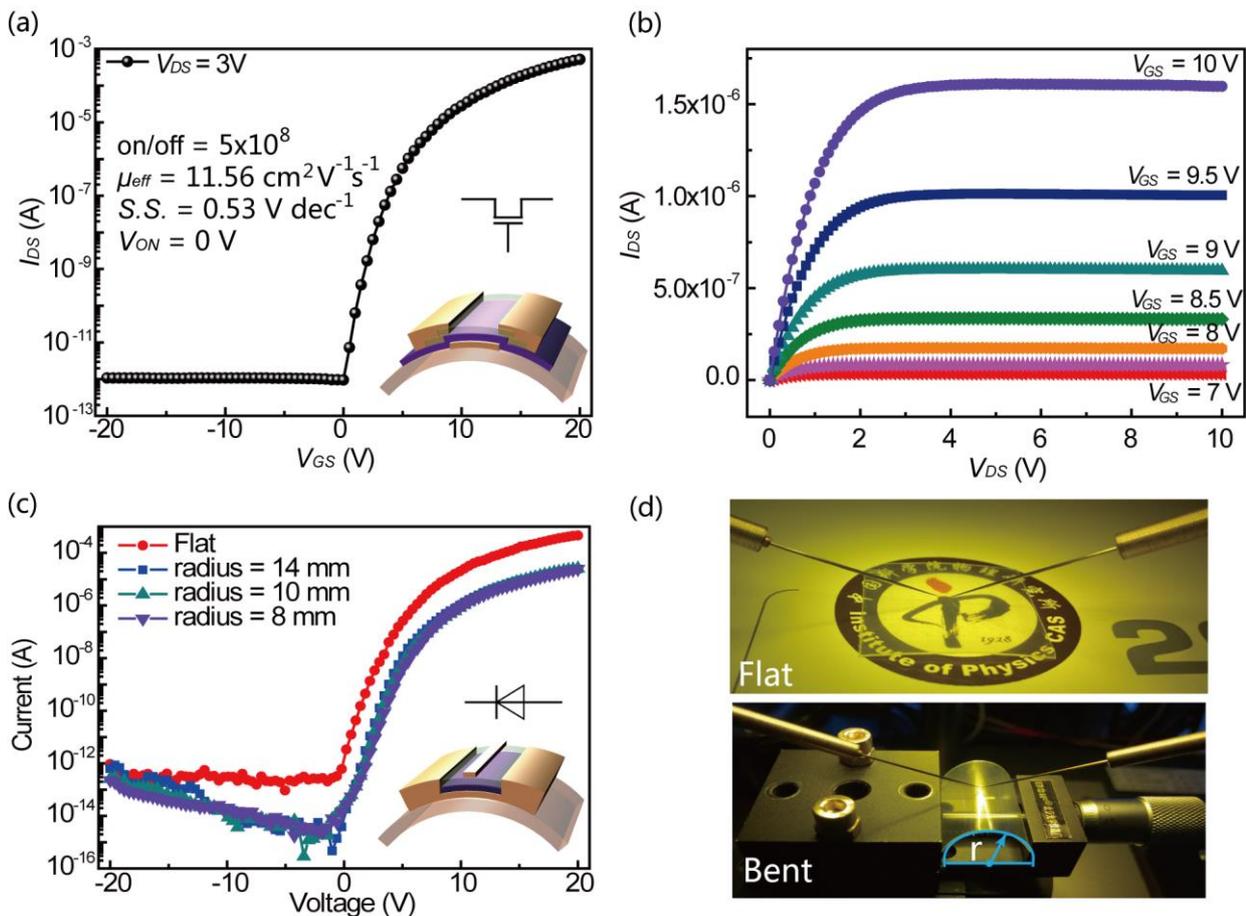

**Figure 2.** Electrical characteristics of referenced TFT and field-effect diode on PEN substrate. (a) Transfer characteristics ($I_{DS}$-$V_{GS}$) and (b) output characteristics ($I_{DS}$-$V_{DS}$) of referenced TFT. (c) Current-voltage (*I-V*) characteristics of field-effect diode with a high rectification ratio of $5 \times 10^8$

while flat and bent. (d) Photograph of devices under test.

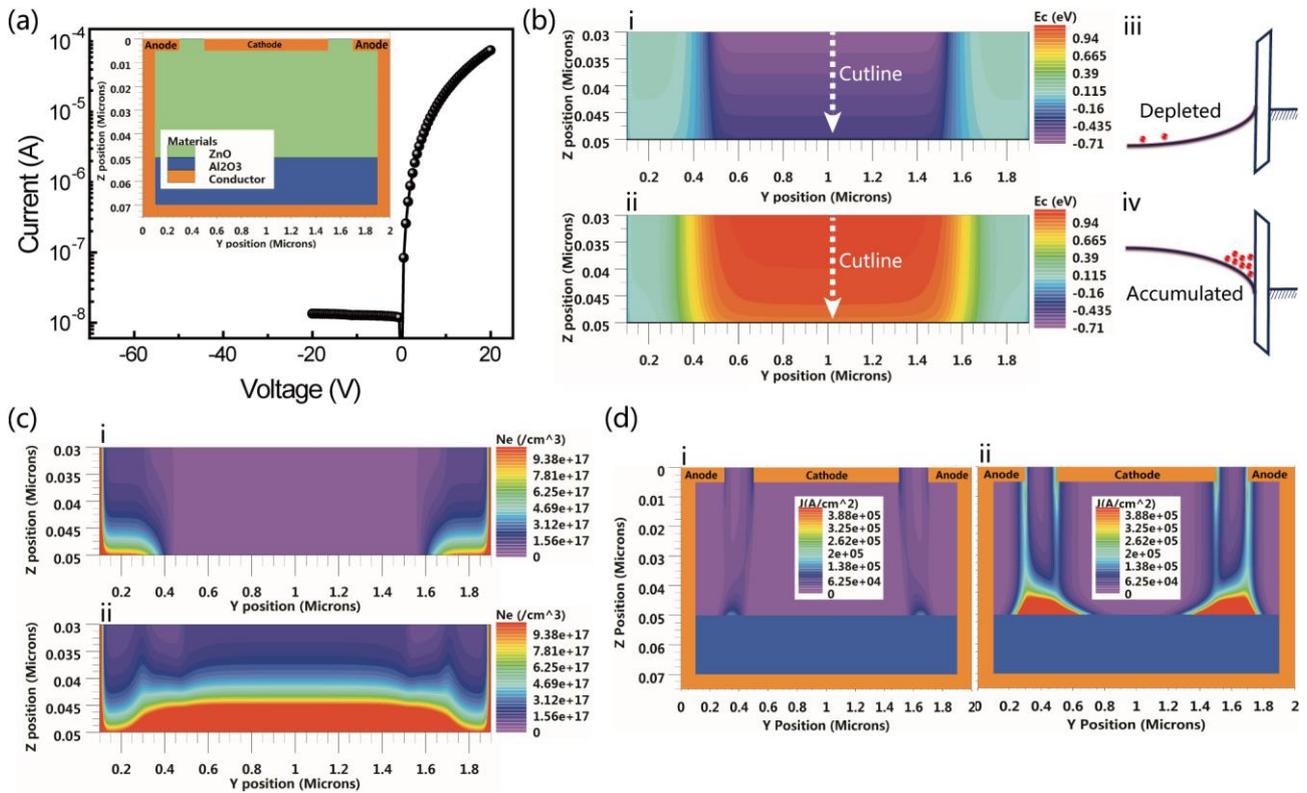

**Figure 3.** Simulation of electrical characteristics of field-effect diode under different voltage biases. (a) Simulated *I-V* characteristics. Inset shows the defined device structure. (b) Conduction band energy (*Ec*) distributions. *Ec* increased along cutline at *V* = -1 V (i), leading to a depleted channel (iii), while decreased at *V* = 1 V (ii), leading to an accumulated channel (iv). (c) Electron concentration (*Ne*) distributions. *Ne* was cut down at *V* = -1 V (i), while boosted up at *V* = 1 V (ii). (d) Current density (*J*) distributions. Negligible current was formed at *V* = -1 V (i), while clear current path formed at *V* = 1 V (ii).

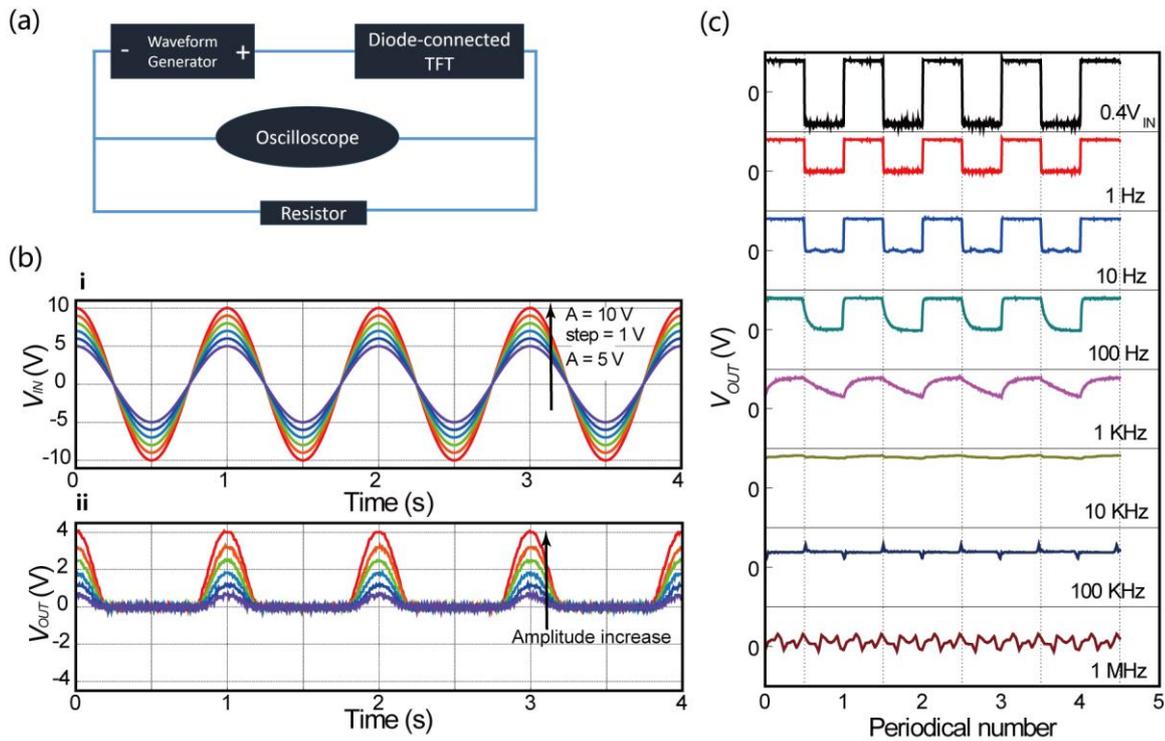

**Figure 4.** Rectification of AC signals. (a) Set-up of single-stage rectifier. (b) Rectification of AC sine signals with different amplitude. (c) Rectification of AC square signals with different frequencies.

**Flexible fully transparent diodes with high rectification ratio of $5 \times 10^8$ are fabricated with all oxide materials at low temperature.** The devices are optically transparent in visible spectra range and electrically robust while mechanically bending. Distinguished from other diodes, these diodes utilize diode-connected thin-film transistor architecture and follow field-effect principles.

**Keyword: diodes, flexible, transparent, oxide, thin-film transistors**

Yonghui Zhang, Zengxia Mei,* Shujuan Cui, Huili Liang, Yaoping Liu and Xiaolong Du

**Flexible Transparent Field-Effect Diodes Fabricated at Low-Temperature with All Oxide Materials**

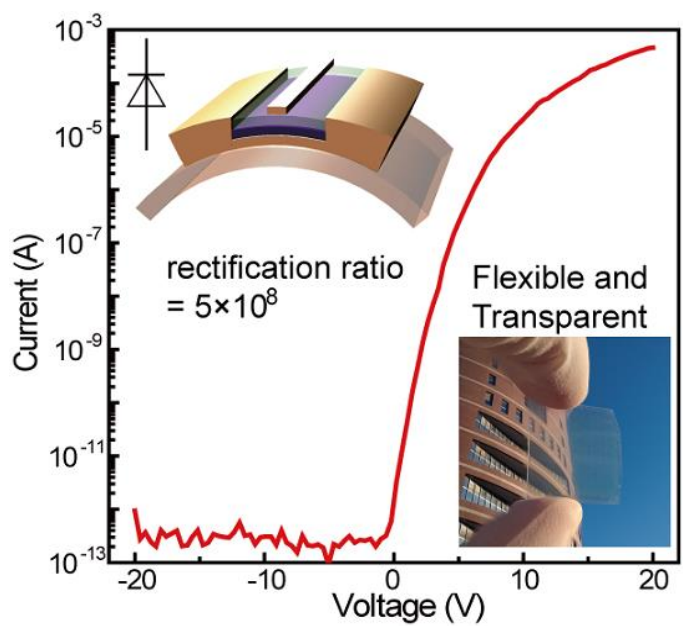